\documentclass[11pt,a4paper]{article}

\usepackage{epsfig,amsmath,amssymb,cite}

\begin{document}

\title{Regular phantom black hole gravitational lensing}
\author{Ernesto F. Eiroa$^{1,2,}$\thanks{e-mail: eiroa@iafe.uba.ar}, Carlos M. Sendra$^{1,2,}$\thanks{e-mail: cmsendra@iafe.uba.ar} \\
{\small $^1$ Instituto de Astronom\'{\i}a y F\'{\i}sica del Espacio,} \\ 
{\small Casilla de Correo 67, Sucursal 28, 1428 Buenos Aires, Argentina}\\
{\small $^2$ Departamento de F\'{\i}sica, Facultad de Ciencias Exactas y 
Naturales, Universidad de Buenos Aires,} \\ 
{\small Ciudad Universitaria Pabell\'on I, 1428 Buenos Aires, Argentina} }

\date{}

\maketitle

\begin{abstract}

We study regular and asymptotically flat phantom black holes as gravitational lenses. We obtain the deflection angle in both the weak and the strong deflection limits, from which we calculate the positions, magnifications, and time delays of the images. We compare our results with those corresponding to the Schwarzschild solution and to the vacuum Brans--Dicke black hole.

\end{abstract}



\section{Introduction}

The subject of black hole gravitational lensing has received growing attention since the discovery of supermassive black holes at the center of galaxies, in particular the one associated with SgrA* in the  Milky Way. The light rays passing close to the photon sphere have a large deviation, and the deflection angle can be calculated by using the strong deflection limit --consisting in a logarithmic approximation-- which was introduced some decades ago \cite{darwin} for the Schwarzschild spacetime. This method allows for the calculation of the positions, the magnifications, and the time delays of the relativistic images. It was rediscovered several times \cite{otros}, then extended to the Reissner--Nordstr\"om metric \cite{eiroto}, and to any spherically symmetric object with a photon sphere \cite{bozza}. 
Numerical studies of black hole lenses were also performed \cite{numerical}. Other interesting works considering strong deflection lenses with spherical symmetry can be found in Refs. \cite{alternative,sdlbd,nakedsing1,nakedsing2,bwlens}. The lensing effects of rotating black holes were analyzed by several researchers \cite{kerr1,kerr2} as well. The apparent shapes or shadows of rotating black holes have a deformation due to the spin \cite{falcke,shadow,kerr2,zakharov,shbw}. It is thought that direct observation of supermassive black holes and the optical effects associated with them will be possible in the near future \cite{zakharov}. For recent reviews about black hole lensing, see Refs. \cite{reviewlens}.

The well known type Ia supernova observations \cite{snova} lead to the cosmological scenario of an accelerated expansion of the Universe. The usual explanation is that the Universe is filled with a negative pressure fluid called dark energy (see, for example, Ref. \cite{de} and references therein) which accounts for about 70 \%, with the other 30 \% corresponding to visible and dark matter. For the prevailing component, the simpler equation of state relating the pressure $p$ with the energy density $\rho$ has the linear form $p=w\rho$: if $w>-1$ it is called quintessence, the case  $w=-1$ corresponds to a cosmological constant $\Lambda$, and when $w<-1$ it receives the name of phantom energy. 
Dark energy can be modeled by a self-interacting scalar field with a potential  \cite{de}. Within this context, regular black hole and wormhole phantom solutions with spherical symmetry were found in Ref. \cite{bronfa}. The stability of these solutions was recently studied \cite{bkz}. 
Static and spherically symmetric solutions with phantom matter, corresponding to regular black holes and black universes --in which an observer gets into an expanding universe after crossing the horizon-- are discussed in Ref. \cite{bdm}. Black holes in scalar-tensor gravity are also analyzed in Ref. \cite{sofa}. Phantom dilaton black holes \cite{gib} were recently studied as gravitational lenses \cite{gyulchev}.

In the present work, we consider as gravitational lenses a class of regular phantom black holes studied in Ref. \cite{bronfa}. In Sec. \ref{defang}, we review the main physical properties of the geometry adopted, introduce the lens equation, and we obtain the exact expression for the deflection angle. In Sec. \ref{wdl}, we approximate the deflection angle by its weak deflection limit value in order to find the positions and magnifications of the primary and secondary images. In Sec. \ref{sdl}, we find the strong deflection limit, from which we calculate the positions and magnifications of the relativistic images. In Sec. \ref{td}, we show the mathematical expressions corresponding to the time delays between the images. Finally, in Sec. \ref{conclu}, we discuss the results obtained and the observational prospects. We use units such that $G=c=1$.

\section{Lens equation and deflection angle}\label{defang}

We consider the Lagrangian corresponding to Einstein gravity coupled to a scalar field $\phi$ with a potential $V(\phi)$ and the electromagnetic field set to zero:
\begin{equation}
L = \sqrt{-g} [R + \epsilon g^{\alpha \beta} \phi_{;\alpha}\phi_{;\beta}- 2 V(\phi)],
\label{lagrangian}
\end{equation}
with $\epsilon =-1$ (phantom field) and 
\begin{equation}
V(\phi) = -\frac{c}{b^2} \left[3 - 2 \cos^2
\left(\frac{\phi}{\sqrt{2}}\right)\right]
- \frac{r_0}{b^3}\left\{3 \sin  \left(\frac{\phi}{\sqrt{2}}\right)\cos \left(\frac{\phi}{\sqrt{2}}  \right)+ \frac{\phi}{\sqrt{2}}\left[ 3 - 2 \cos^2 \left(\frac{\phi}{\sqrt{2}}\right)\right] \right\}. 
\end{equation}
The Einstein-scalar equations coming from this Lagrangian admit a static and spherically symmetric solution \cite{bronfa} having the metric
\begin{equation}
ds^{2}=-A(r)dt^{2}+B(r)dr^{2}+C(r)(d\theta^{2}+\sin^{2}\theta d\phi^{2}),
\label{m1}
\end{equation}
with
\begin{equation}
A(r)=B(r)^{-1}=1+\frac{r_0 r}{b^2}+(r^2+b^2)\left[\frac{c}{b^2}+\frac{r_0}{b^3}\tan^{-1}\left(\frac{r}{b}\right)\right],  
\hspace{1cm} 
C(r)=r^2+b^2,
\label{m2}
\end{equation}
and the scalar field
\begin{equation}
\phi = \sqrt{2} \tan^{-1} \left(\frac{r}{b}\right),
\label{phi1}
\end{equation}
where $c$, $r_0$, and $b>0$ are constants. The parameter $b$ can be interpreted as a scale of the scalar field. The radial coordinate $r$ is a real number, and the function $\mathcal{S}(r)=4 \pi R^2(r)$, with $R(r)=\sqrt{C(r)}$, gives the area of the spherical surface corresponding to a given value $r$. This area  function $\mathcal{S} (r)$ has the minimum value $\mathcal{S}(0)=4 \pi b^2$, so the geometry presents a throat at $r_{th}=0$. The solution is regular everywhere and the position of the horizon can be obtained from the equation $A(r)=0$. The metric becomes asymptotically flat \cite{bronfa} as $r \rightarrow +\infty$ when
\begin{equation}
c=-\frac{\pi r_0}{2b}.
\label{c}
\end{equation}
The constant $m=r_0/3$ can be interpreted in the usual way as the mass. If $m=0$ the geometry is that of the Ellis wormhole \cite{ellis}, which connects two symmetric asymptotically flat regions through a throat at $r_{th}=0$. When $m<0$ there are no horizons and the spacetime corresponds to a wormhole with a throat at $r_{th}=0$ that connects an asymptotically flat region ($r>0$) with an asymptotically anti-de Sitter region ($r<0$). 
If $m>0$, there is one Killing horizon, corresponding to the only root $r_h$ of $A(r)$. In this case, the region corresponding to $r>r_h$ is asymptotically flat and the one with $r<r_h$ is asymptotically de Sitter. These regular black holes also receive the name of black universes. When $0<b< 3 \pi m /2$ one has $0<r_h<2m$, if $b= 3 \pi m /2$ then $r_h=0$, and when $b> 3 \pi m /2$ one obtains $r_h<0$. In the first case $r_{th}=0<r_h$, and the throat is not a true one, because $r$ corresponds to a time coordinate in the region with $r<r_h$; in the second case, the throat and the horizon coincides ($r_{th}=r_h=0$); and in the third case, the throat is outside the horizon ($r_h<r_{th}=0$). The regular phantom solution combines the properties of black holes (the presence of a horizon) with those of wormholes (the presence of a throat). In the limit $b\rightarrow 0$ the throat is lost and for $m>0$ the Schwarzschild solution, with the singularity at $r=0$ and the horizon at $r_h=2m$, is obtained. For more 
details 
see Refs. \cite{bronfa,bkz}.

\begin{figure}[t!]
\begin{center}
    \includegraphics[scale=0.9,clip=true,angle=0]{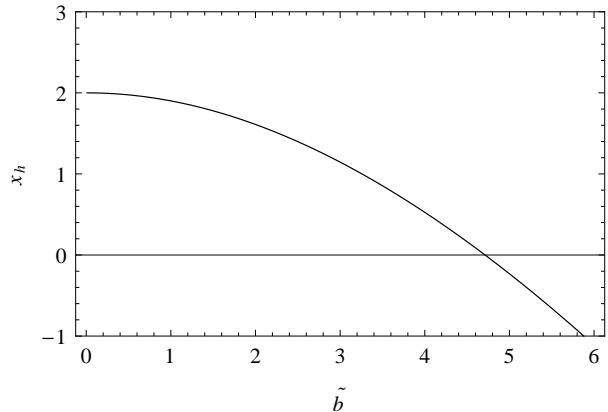}
    \caption{Adimensionalized radial coordinate $x_h=r_h/m$ of the horizon as a function of the parameter $\tilde{b}=b/m$. When $\tilde{b}> 3 \pi /2$  the horizon radial coordinate is negative and there is a throat outside the horizon, located at $x_{th}=r_{th}/m=0$.}
    \label{horizonfig}
\end{center}
\end{figure}

 From now on, we adopt the value of $c$ given by the condition (\ref{c}), so that the geometry is asymptotically flat, and $m>0$ corresponding to a black hole (or black universe). It is useful for the calculations that follow to adimensionalize all quantities in terms of the mass $m$, by defining the radial coordinate $x=r/m$, the time coordinate $T=t/m$, and the parameter $\tilde{b}=b/m$. Then the solution takes the form
\begin{equation}
ds^{2}=-A(x)dT^{2}+B(x)dx^{2}+C(x)(d\theta^{2}+\sin^{2}\theta d\phi^{2}),
\label{m3}
\end{equation}
with
\begin{equation}
A(x)=B(x)^{-1}=1+\frac{3x}{\tilde{b}^2}+\frac{3}{\tilde{b}}\left(1+\frac{x^2}{\tilde{b}^2}\right)\left[-\frac{\pi}{2}+\tan^{-1}\left( \frac{x}{\tilde{b}}\right)\right],  
\hspace{1cm} 
C(x)=x^2+\tilde{b}^2,
\label{m4}
\end{equation}
and
\begin{equation}
\phi = \sqrt{2} \tan^{-1} \left(\frac{x}{\tilde{b}}\right).
\label{phi2}
\end{equation}
The value of the radial coordinate corresponding to the horizon $x_h$, obtained numerically from the condition $A(x_h)=0$, is the decreasing function of $\tilde{b}$ shown in Fig. \ref{horizonfig}; for $\tilde{b}=3\pi/2$ the horizon radial position is $x_h=0$ and when $\tilde{b}>3\pi/2$ it is negative. The geometry has a photon sphere, its radius $x_{ps}$ is given by the largest positive solution of the equation
\begin{equation}
\frac{A'(x)}{A(x)}=\frac{C'(x)}{C(x)},
\label{xps}
\end{equation}
where the prime represents the derivative with respect to $x$. Replacing the metric functions in  Eq. (\ref{xps}) and after some straightforward calculations, we find that the photon sphere radius has the constant value $x_{ps}=3$.

\begin{figure}[t!]
\begin{center}
    \includegraphics[scale=0.9,clip=true,angle=0]{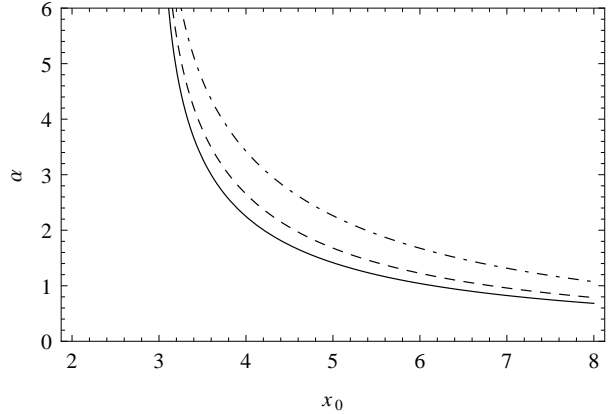}
    \caption{Deflection angle as a function of the adimensionalized closest approach distance $x_0=r_0/m$, for three representative values of the parameter $\tilde{b}$: $1$ (solid line), $3$ (dashed line), and $6$ (dashed-dotted line). The deflection angle diverges when $x_0=x_{ps}=3$ (photon sphere).}
    \label{adex}
\end{center}
\end{figure}

To study the lensing effects, we adopt the configuration where the black hole ($l$) is situated between a point source of light ($s$) and an observer ($o$), both of them located in the region corresponding to $x>0$, at distances much larger than the horizon radius $x_h$, so that they lie in a flat region. The lens equation relates the deflection angle $\alpha$ with the angular positions --seen from the observer, with the optical axis defined as the line joining the observer and the lens-- of the source $\beta$ and the images $\theta$.  We will take $\beta >0$ without losing generality. The lens equation can be written in the form \cite{lenseq}
\begin{equation}
\tan \beta =\frac{d_{ol}\sin \theta - d_{ls} \sin (\alpha -\theta)}{d_{os} \cos (\alpha -\theta)} ,
\label{pm1}
\end{equation}
where  $d_{os}$, $d_{ol}$, and $d_{ls}$, are the observer--source, observer--lens, and lens--source angular diameter (adimensionalized) distances, respectively. The deflection angle for a photon coming from infinity, in terms of the closest approach distance $x_0$, is given by \cite{nakedsing1,weinberg}
\begin{equation}
\alpha(x_0)=I(x_0)-\pi,
\label{alpha1}
\end{equation}
where
\begin{equation}
I(x_0)=\int^{\infty}_{x_0}\frac{2\sqrt{B(x)}dx}{\sqrt{C(x)}\sqrt{A(x_0)C(x)\left[ A(x)C(x_0)\right]^{-1}-1}}.
\label{i0}
\end{equation}
The deflection angle is a monotonic decreasing function of $x_0$; it diverges when $x_0$ gets close to the radius of the photon sphere $x_{ps}$ and it approaches to zero for large $x_0$, as it can be seen in Fig. \ref{adex}. When $x_{0}$ is close enough to $x_{\mathrm{ ps }}$, the deflection angle  $\alpha $ is greater than $2\pi $, and the photons perform one or more turns around the black hole before reaching the observer. Then, there are two infinite sets of strong deflection or relativistic images, one of them due to clockwise winding around the black hole and the other one produced by counterclockwise winding. These relativistic images are located, respectively, at the same side and at the opposite side of the source. 
To obtain the positions of the images for a given source position, we have to replace the deflection angle given by  Eqs. (\ref{alpha1}) and (\ref{i0}) in the lens equation (\ref{pm1}) and invert it. The resulting equation is quite complicated and cannot be solved analytically without some simplifying approximations.

\section{Weak deflection images}\label{wdl}

Let us first analyze the case of photons with a large impact parameter, so the closest approach distance $x_0$ is large. By defining $y=x_0/x$, and performing a Taylor expansion to second order around $1/x_0$ of the integrand, the Eq. (\ref{i0}) takes the form
\begin{equation}
I(x_0)=\int^{1}_{0} f(y)dy,
\label{idl}
\end{equation}
where
\begin{equation}
f(y)\approx\frac{2}{\sqrt{1-y^2}}+\frac{2(1+y+y^2)}{(1+y)\sqrt{1-y^2}}\frac{1}{x_0}+\frac{-b^2(-1+y)(1+y)^3+3(1+y+y^2)^2}{y(1+y)^2\sqrt{-1+y^{-2}}}\frac{1}{x^{2}_{0}};
\end{equation}
by calculating the integral and replacing it in Eq. (\ref{alpha1}), the deflection angle, in the weak deflection limit, results
\begin{equation}
\alpha(x_0)\approx \frac{4}{x_0}+\frac{-16+(15+\tilde{b}^2)\pi}{4x^{2}_{0}}.
\label{alphawdl}
\end{equation}
Keeping only the first order of Eq. (\ref{alphawdl}), and using the relation $x_0=d_{ol}\sin\theta\approx d_{ol}\theta$, the deflection angle finally is
\begin{equation}
\alpha(\theta)\approx\frac{4}{d_{ol}}\frac{1}{\theta}.
\label{alfawd3}
\end{equation}
For high alignment, the lens equation (\ref{pm1}) can be written in a simpler form, by approximating the trigonometric functions by their first order expansions, so it reduces to
\begin{equation}
\beta=\theta-\frac{d_{ls}}{d_{os}}\alpha .
\label{lenseq}
\end{equation}
The Einstein ring is formed for perfect alignment of the source, the lens, and the observer, i.e. when $\beta=0$; its radius is given by
\begin{equation}
\theta_{E}=\sqrt{\frac{4d_{ls}}{d_{ol}d_{os}}}.
\label{einstein1}
\end{equation}
Then, in terms of the Einstein radius, the deflection angle has the form
\begin{equation}
\alpha(\theta)\approx\frac{\theta^{2}_{E}d_{os}}{d_{ls}}\frac{1}{\theta}.
\label{alfaw1}
\end{equation}
The angular positions of the primary and the secondary images are obtained by replacing (\ref{alfaw1}) in the lens equation (\ref{lenseq}) to give
\begin{equation}
\theta_{\rm{p,s}}=\frac{\beta\pm\sqrt{\beta^2+4\theta^{2}_{E}}}{2}.
\label{thetapm}
\end{equation}

Gravitational lensing conserves surface brightness. The magnification is given by the quotient of the solid angles subtended by the image and the source
\begin{equation}
\mu =\left|\frac{\sin\beta}{\sin\theta}\frac{d\beta}{d\theta}\right|^{-1},
\label{magnif1}
\end{equation}
which for small angles reduces to
\begin{equation}
\mu =\left|\frac{\beta}{\theta}\frac{d\beta}{d\theta}\right|^{-1}.
\label{magnif2}
\end{equation}
By replacing the angular positions of the images given by (\ref{thetapm}) in the expression (\ref{magnif2}), the magnifications of the primary and secondary images take the form
\begin{equation}
\mu_{\rm{p,s}}=\frac{\left(\beta \pm \sqrt{\beta^2+4\theta^{2}_{E}}\right)^2}{4\beta\sqrt{\beta^2+4\theta^{2}_{E}}}.
\label{magpm}
\end{equation}

It is important to note that the results obtained above do not depend on $\tilde{b}$ to first order in $1/x_0$. Higher order corrections (which are functions of $\tilde{b}$) can be obtained following the procedure detailed in Ref. \cite{kepe}.

\section{Relativistic images}\label{sdl}

\begin{figure}[t!]
\begin{center}
    \includegraphics[scale=0.9,clip=true,angle=0]{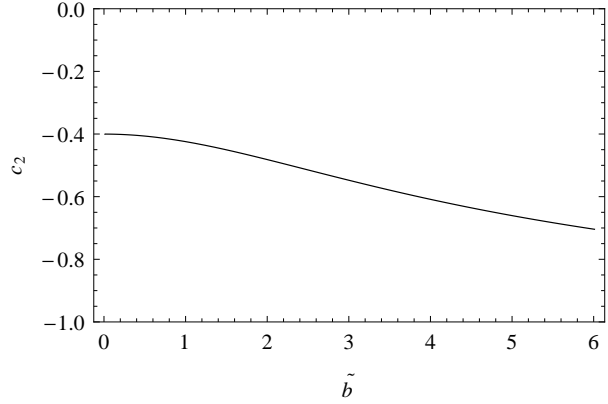}
    \caption{Strong deflection limit coefficients $c_1$ and $c_2$ as functions of the parameter $\tilde{b}$: the coefficient $c_1=1$ is a constant and $c_2$ is the decreasing function shown in the plot.}
    \label{c1yc2}
\end{center}
\end{figure}

We consider now the case of photons passing close to the photon sphere, for which we adopt the so-called strong deflection limit \cite{bozza}. We split the integral (\ref{i0}) as a sum of two parts:
\begin{equation}
I(x_0)=I_D(x_0)+I_R(x_0),
\label{i0n}
\end{equation}
where
\begin{equation}
I_D(x_0)=\int^{1}_{0}R(0,x_{ps})f_0(z,x_0)dz
\label{id}
\end{equation}
and
\begin{equation}
I_R(x_0)=\int^{1}_{0}[R(z,x_0)f(z,x_0)-R(0,x_{ps})f_0(z,x_0)]dz,
\label{ir}
\end{equation}
with
\begin{equation}
z=\frac{A(x)-A(x_0)}{1-A(x_0)},
\label{z}
\end{equation}
\begin{equation}
R(z,x_0)=\frac{2\sqrt{A(x)B(x)}}{A'(x)C(x)}[1-A(x_0)]\sqrt{C(x_0)},
\label{r}
\end{equation}
\begin{equation}
f(z,x_0)=\frac{1}{\sqrt{A(x_0)-[(1-A(x_0))z+A(x_0)]C(x_0)[C(x)]^{-1}}}.
\label{f}
\end{equation}
By performing a Taylor expansion of the argument inside the square root in Eq. (\ref{f}), one has
\begin{equation}
f_0(z,x_0)=\frac{1}{\sqrt{\varphi (x_0) z+\gamma (x_0) z^{2}}},
\label{f0}
\end{equation}
where
\begin{equation}
\varphi (x_0)=\frac{1-A(x_0)}{A'(x_0) C(x_0)}\left[ A(x_0) C'(x_0) - A'(x_0) C(x_0)\right],
\label{varphi}
\end{equation}
and
\begin{eqnarray}
\gamma (x_0) &=& \frac{\left[ 1-A(x_0)\right] ^{2}}{2[A'(x_0)]^{3} [C(x_0)]^{2}}\left\{ 2 [A'(x_0)]^{2} C(x_0) C'(x_0) - A(x_0) A''(x_0) C(x_0) C'(x_0) \right. \nonumber \\
&& \left. + A(x_0) A'(x_0) \left[ C(x_0) C''(x_0) -2 [C'(x_0)]^{2}\right] \right\} .
\label{gamma}
\end{eqnarray}
With these definitions, $I_D$ converges for $x_0\neq x_{ps}$ because $\varphi \neq 0$ and $f_0\sim 1/\sqrt{z}$. If $x_0=x_{ps}$, from Eq. (\ref{varphi}), we find that $\varphi=0$ and $f_0\sim1/z$, and $I_D(x_0)$ has a logarithmic divergence. So, $I_D$ is the term containing the divergence at $x_0=x_{ps}$, and $I_R$ is regular since it has the divergence subtracted. The logarithmic divergence of the deflection angle for photons passing close to the photon sphere, in terms of the impact parameter
\begin{equation}
u=\sqrt{\frac{C(x_0)}{A(x_0)}},
\label{u}
\end{equation}
can be approximated by the simple general form \cite{bozza} 
\begin{equation}
\alpha(u)=-c_1\ln\left(\frac{u}{u_{ps}}-1\right)+c_2+O(u-u_{ps}),
\label{alfasdl}
\end{equation}
where $u_{ps}$ is the impact parameter evaluated at $x_0=x_{ps}$ and
\begin{equation}
c_1=\frac{R(0,x_{ps})}{2\sqrt{\gamma(x_{ps})}}
\label{c1}
\end{equation}
and
\begin{equation}
c_2=-\pi+c_R+c_1\ln \frac{2\gamma(x_{ps})}{A(x_{ps})},
\label{c2}
\end{equation}
with
\begin{equation}
c_R=I_R(x_{ps}).
\label{cr}
\end{equation}
The quantities $c_1$ and $c_2$ are named the strong deflection limit coefficients, which depend only on the metric functions. For the phantom black hole, we obtain that the critical impact parameter is given by 
\begin{equation}
u_{ps}=\sqrt{\frac{2\tilde{b}^3}{2\tilde{b}-3\pi+6\tan^{-1}\left(\frac{3}{\tilde{b}}\right)}},
\end{equation}
and the strong deflection limit coefficients have the values
\begin{equation}
c_1=1,
\label{c1bh}
\end{equation}
and
\begin{equation}
c_2=-\pi+c_R+c_1\ln \frac{\tilde{b}^3 \left[-6 \tilde{b} + 9 \pi + \tilde{b}^2 \pi - 2 (9 + \tilde{b}^2) \tan^{-1}\left(\frac{3}{\tilde{b}}\right)\right]^2}{(9 +
   \tilde{b}^2)^2 \left[2 \tilde{b} - 3 \pi + 6 \tan^{-1}\left(\frac{3}{\tilde{b}}\right)\right]^3},
\label{c22}
\end{equation}
where $c_R$ cannot be calculated analytically, so it is found numerically for each value of $\tilde{b}$. The coefficient $c_1$ is a constant and $c_2$ is shown as a function of the parameter $\tilde{b}$ in Fig. \ref{c1yc2}.

The deflection angle, obtained above in the strong deflection limit, can be directly related to the positions and magnifications of the relativistic images by using the lens equation. The lensing effects are more significant when $\beta$ and $\theta$ are small, i.e. when the objects are highly aligned. In  this case, $\alpha$ is close to an even multiple of $\pi$. For $\beta\neq 0$, two infinite sets of relativistic images are obtained, one on each side of the lens. So, the deflection angle for the first set of relativistic images can be written as $\alpha=2n\pi+\Delta\alpha_n$, with $n \in \mathbb{N}$, and $0<\Delta\alpha_n\ll 1$. In this approximation, the lens equation (\ref{pm1}) takes the simple form
\begin{equation}
\beta =\theta -\frac{d_{ls}}{d_{os}}\Delta \alpha _{n}.
\label{pm2}
\end{equation}
For the other set of images, which satisfy $\alpha =-2n\pi -\Delta \alpha _{n}$, the quantity $\Delta \alpha _{n}$ is replaced by $-\Delta \alpha _{n}$ in Eq. (\ref{pm2}). The deflection angle can also be expressed in terms of two measurable magnitudes: the angular position of the image $\theta$ and the distance between the observer and the black hole $d_{ol}$. According to the lens geometry, we have that $u=d_{ol}\sin\theta\approx d_{ol}\theta$, so Eq. (\ref{alfasdl}) results in
\begin{equation}
\alpha (\theta )\approx-c_{1}\ln \left( \frac{d_{ol}\theta }{u_{ps}}-1 \right)
+c_{2}.
\label{pm4}
\end{equation}
The angular position of the $n$-th image is obtained by inverting Eq. (\ref{pm4}) and performing a first order Taylor expansion around $\alpha=2n\pi$:
\begin{equation}
\theta _{n}=\theta ^{0}_{n}-\zeta _{n}\Delta \alpha _{n},
\label{pm6}
\end{equation}
where
\begin{equation}
\theta ^{0}_{n}=\frac{u_{ps}}{d_{ol}}\left[ 1+e^{(c_{2}-2n\pi )/c_{1}}
 \right] ,
\label{pm7}
\end{equation}
and
\begin{equation}
\zeta _{n}=\frac{u_{ps}}{c_{1}d_{ol}}e^{(c_{2}-2n\pi )/c_{1}}.
\label{pm8}
\end{equation}
From Eqs. (\ref{pm2}) and (\ref{pm6}), $\theta_n$ can be rewritten using $\Delta\alpha_n=(\theta_n-\beta)d_{ol}/d_{ls}$. Then,
\begin{equation}
\theta _{n}=\theta ^{0}_{n}-\frac{\zeta _{n}d_{os}}{d_{ls}}(\theta _{n}-\beta ).
\label{pm10}
\end{equation}
Since $0<\zeta_n d_{os}/d_{ls}<1$ and keeping only the first-order term in $\zeta_n d_{os}/d_{ls}$, which is a small correction to $\theta^{0}_{n}$, the angular positions for one set of relativistic images finally take the form
\begin{equation}
\theta _{n}=\theta ^{0}_{n}+\frac {\zeta _{n}d_{os}}{d_{ls}}(\beta -\theta ^{0}_{n}),
\label{pm14}
\end{equation}
and for the other set
\begin{equation}
\theta _{n}=-\theta ^{0}_{n}+\frac {\zeta _{n}d_{os}}{d_{ls}}(\beta +\theta ^{0}_{n}).
\label{pm15}
\end{equation}

The magnification $\mu_n$ of the $n$-th relativistic image is given by the angle subtended by the image and the source, as in the weak deflection case, i.e. Eq. (\ref{magnif1}). Considering small angles and replacing Eq. (\ref{pm14}) in the  expression (\ref{magnif2}), we have
\begin{equation}
\mu _{n}=\frac{1}{\beta}\left[ \theta ^{0}_{n}+
\frac {\zeta _{n}d_{os}}{d_{ls}}(\beta - \theta ^{0}_{n})\right]
\frac {\zeta _{n}d_{os}}{d_{ls}}.
\label{pm18}
\end{equation}
After performing a Taylor expansion in $\zeta_n d_{os}/d_{ls}$, the magnification of the $n$-th image for both sets of relativistic images finally reduces to
\begin{equation}
\mu _{n}=\frac{1}{\beta}\frac{\theta ^{0}_{n}\zeta _{n}d_{os}}{d_{ls}}.
\label{pm19}
\end{equation}
Equations (\ref{pm7}) and (\ref{pm8}) imply that the magnifications decrease exponentially with $n$, so the first relativistic image is the brightest one. On the other hand, the factor $(u_{ps}/d_{ol})^2$ is very small meaning that the magnifications are very faint unless the lens and the source are highly aligned ($\beta\approx 0$).

\begin{figure}[t!]
\begin{center}
    \includegraphics[scale=0.9,clip=true,angle=0]{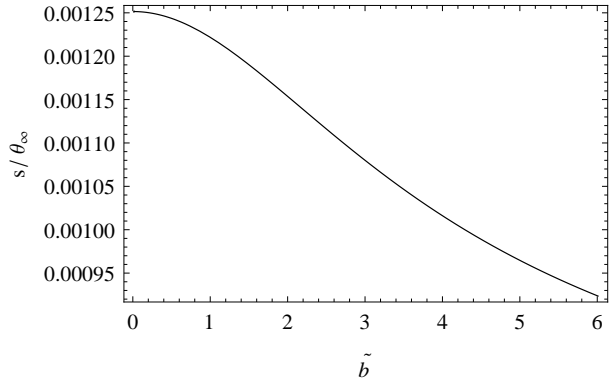}
    \caption{The observables $s/\theta_{\infty}$ and $r$: the quotient $s/\theta_{\infty}$ is the decreasing function of $\tilde{b}$ shown in the plot, and $r=e^{2\pi}$ is a constant.}
    \label{observables}
\end{center}
\end{figure}

These results can be compared with observations by defining the observables \cite{bozza}:
\begin{equation}
\theta_{\infty}=\frac{u_{ps}}{d_{ol}},
\label{ob1}
\end{equation}
\begin{equation}
s=\theta_1-\theta_{\infty},
\label{ob2}
\end{equation}
and
\begin{equation}
r=\frac{\mu_1}{\sum_{n=2}^{\infty}\mu_n}.
\label{ob3}
\end{equation}
The limiting value $\theta_{\infty}$, where the images approach as $n\rightarrow \infty$, is an increasing function of $\tilde{b}$ for a given value of $d_{ol}$. As the first relativistic image is the outermost and brightest one, it would be resolved from the others, so $s$ is defined as the angular separation between the first relativistic image and the others, which approach to the limiting angular position $\theta_{\infty}$. The observable $r$ is the quotient between the flux of the first image and the flux coming from all the other ones. As it is shown in the above expressions, the angular positions and the magnifications of the relativistic images are related to the strong deflection limit coefficients. For high alignment, the observables take the form
\begin{equation}
s=\theta_{\infty}e^{(c_2-2\pi)/c_1}=\theta_{\infty}e^{c_2-2\pi},
\label{ob4}
\end{equation}
and
\begin{equation}
r=e^{2\pi/c_1}=e^{2\pi}.
\label{ob5}
\end{equation}
The observable $r$ is a constant and the quotient $s/\theta_{\infty}$ is plotted as a function of the parameter $\tilde{b}$ in Fig. \ref{observables}.

\section{Time delays}\label{td} 

In the case of transient sources, it is of interest to study the time delays between the images, due to the different paths that follow the photons that form them. The (adimensionalized) time delay between the primary and the secondary images is given by \cite{schneider}
\begin{equation}
\Delta T_{\rm{p,s}}=4\left( \frac{\theta_{\rm{s}}^{2}-\theta_{\rm{p}}^{2}}{2|\theta_{\rm{p}}\theta_{\rm{s}}|} +\ln \left|\frac{\theta_{\rm{s}}}{\theta_{\rm{p}}}\right|\right),
\label{td1}
\end{equation}
which can be written in the form
\begin{equation}
\Delta T_{\rm{p,s}}=4\left( \frac{-\beta \sqrt{\beta^{2}+4\theta_{\rm{E}}^{2}}}{2\theta_{\rm{E}}^{2}}+\ln \left| \frac{\beta -\sqrt{\beta^{2}+4\theta_{\rm{E}}^{2}}}{\beta +\sqrt{\beta^{2}+4\theta_{\rm{E}}^{2}}}\right|\right).
\label{td2}
\end{equation}
From Eq. (\ref{td2}), it is clear that for perfect alignment (i.e. $\beta=0$), there is no time delay. As $\beta /\theta_{\rm{E}}$ grows, larger time delays can be obtained, but if $\beta /\theta_{\rm{E}}\gg 1$, the magnification of the primary image is close to one and the secondary image is very faint and it could not be observable. The optimal situation is when $\beta /\theta_{\rm{E}}$ is small enough to have large magnifications of both images, but not too close to zero, so the time delay can be longer than the typical time scale of the variable source. By using the results of Sec. \ref{wdl}, we see that, in the first order approximation adopted in the present work, the value of $\Delta T_{\rm{p,s}}$ is the same as for the Schwarzschild black hole.  \\

For the relativistic images, the (adimensionalized) time delay between the \textit{n}-th and \textit{m}-th images formed at the same side of the lens is given by \cite{bozmanc}
\begin{eqnarray}
\Delta T^{\rm{s}}_{n,m} &=& u_{\rm{ps}}\left\{ 2\pi (n-m)+2\sqrt{2}\left[ e^{(c_{2}-2m\pi)/(2c_{1})}-e^{(c_{2}-2n\pi)/(2c_{1})}\right] \right.
\nonumber \\
&& \left. \pm \frac{\sqrt{2}d_{\rm{os}}\beta}{c_{1}d_{\rm{ls}}}\left[ e^{(c_{2}-2m\pi)/(2c_{1})}-e^{(c_{2}-2n\pi)/(2c_{1})}\right] \right\} ,
\label{td3}
\end{eqnarray}
where the upper (lower) sign corresponds if both images are on the same (opposite) side of the source. For the images at the opposite side of the lens we have \cite{bozmanc}
\begin{eqnarray}
\Delta T^{\rm{o}}_{n,m} &=& u_{\rm{ps}}\left\{ 2\pi (n-m) +2\sqrt{2}\left[ e^{(c_{2}-2m\pi)/(2c_{1})}-e^{(c_{2}-2n\pi)/(2c_{1})}\right] \right.
\nonumber \\
&& \left. +\frac{\sqrt{2}d_{\rm{os}}\beta}{c_{1}d_{\rm{ls}}}\left[ e^{(c_{2}-2m\pi)/(2c_{1})}+e^{(c_{2}-2n\pi)/(2c_{1})}\right]-\frac{2d_{os}\beta}{d_{ls}}\right\},
\label{td4}
\end{eqnarray}
where the image with winding number $n$ is on the same side of the source and the other one on the opposite side. In both cases, the expressions from Ref. \cite{bozmanc} have been expanded to first order in the source position angle, measured from the observer instead of from the source. The first term in Eqs. (\ref{td3}) and (\ref{td4}) is by large the most important one \cite{bozmanc}, and is the delay related to the difference of loops given by the photons around the black hole before emerging. The time delay between the primary and the secondary images is shorter than the time delays between the relativistic images.

\section{Conclusions}\label{conclu}

In this paper, we have studied the gravitational lensing effects produced by regular black holes, which are solutions of the Einstein equations with a scalar field possessing a negative kinetic term and a potential, without the presence of the electromagnetic field. These black holes (or black universes) have an asymptotically flat region that continues to an asymptotically de Sitter region after crossing the horizon, and they can have a throat outside the horizon. We have obtained the weak and the strong deflection limits of the deflection angle in order to calculate the positions, magnifications and time delays of the images for a high alignment scenario. The results were shown in terms of the quotient between the parameter $b$ associated to the ghost field and the (positive) mass $m$ of the black hole. 

Let us compare the results obtained in the present work, with those corresponding to the Schwarzschild black hole solution. In the weak deflection limit, to first order in the quotient $m/r_0$ between the mass and the closest approach distance $r_0$, we have found that the deflection angle does not depend on $b/m$ and the positions, magnifications and time delays of the images are the same as in the case of the Schwarzschild spacetime. The strong deflection limit coefficients for the Schwarzschild black holes \cite{bozza} have the values  $c_1^{Sch}=1$ and $c_2^{Sch}=\ln[216(7-4\sqrt{3})] - \pi \approx -0.400230$. For the phantom black holes, we have obtained that $c_1=c_1^{Sch}$, and that $c_2$ is a decreasing function of $b/m$, which is equal to $c_2^{Sch}$ in the limit $b\rightarrow 0$. For a given value of the distance $D_{ol}$ between the observer and the black hole, we have found that the limiting value of the angular position of the images $\theta_\infty$ is larger than $\theta_\infty ^{Sch}=3\sqrt{3} \, m/D_{ol}\approx 5.19615 \, m/D_{ol}$, and the relative separation of the images
$s/\theta_\infty$ is smaller than $(s/\theta_\infty)^{Sch}= 0.00125$, so the images are farther from the origin and more packed together than in the Schwarzschild case. With respect to the magnifications, the behavior is similar to that of the Schwarzschild lenses, in the sense that the quotient between the flux of the first image and the flux coming from all the others satisfy $r=r^{Sch}=e^{2\pi}$. The complicated expressions for the time delays of the relativistic images makes a general comparison difficult, so we have to rely on a numerical example (see below). One particularly interesting case occurs if $b/m=3 \pi/2$, i.e. when the horizon coincides with the throat, since it was shown in Ref. \cite{bkz} that the solution is stable; we have that $c_2=-0.646528$, $\theta_\infty =7.84411 \, m/D_{ol}$ and $s/\theta_\infty = 0.000978$, with large differences with respect to the Schwarzschild values.

To provide a numerical example, let us consider the Galactic center supermassive black hole \cite{guillessen}, for which the mass is $M=4.31 \times 10^{6}M_{\odot}$  and the distance from the Earth is $D_{ol}=8.33$ kpc. We also adopt $D_{os}=2D_{ol}$ as the value of the distance between the observer and the source, an angular position of the source $\beta=0.5 \, \theta_\infty$, and $b/m=3 \pi/2$. Then, we have that the limiting value of the angular positions of the relativistic images is $\theta_\infty=40.0183$ $\mu$as, with the first image separated from it by $s=0.03915$ $\mu$as. The magnification of the first strong deflection image is $\mu_1=7.6\times 10^{-13}$, and the quotient between the flux of the first image and the flux coming from all the others is $r=e^{2\pi}\approx 535$. The time delay between the first relativistic image at one side and the first one at the other side is $|\Delta t^{\rm{o}}_{1,1}|=1.122\times10^{-9}$ min, and the time delay between the first relativistic image and the second one at the same side is $|\Delta t^{\rm{s}}_{1,2}|=17.638$ min. The corresponding values for the Schwarzschild black hole are $\theta_\infty^{Sch}=26.5093$ $\mu$as, $s^{Sch}=0.03318$ $\mu$as, $\mu_1^{Sch}=6.4\times 10^{-13}$, $r^{Sch}=r$, $|\Delta t^{\mathrm{o}\;\;\; Sch}_{1,1}|=4.952 \times 10^{-10}$ min, and $|\Delta {t^{\mathrm{s}\;\;\; Sch}_{1,2}}|=11.704$ min. We see that for the phantom black hole the time delays are, in both cases, larger than those for the Schwarzschild geometry.

Another interesting spacetime for a comparison with our results is the spherically symmetric vacuum solution of Brans-Dicke theory, for which the gravitational lensing in the strong deflection limit was studied in Ref. \cite{sdlbd}. The theory has a coupling parameter $\omega$; in the limit $\omega\rightarrow \infty$ the Schwarzschild geometry is recovered. Following Ref. \cite{sdlbd}, we adopt for the calculations the values $\omega=500$ and $\omega=50000$. The strong deflection limit coefficients \cite{sdlbd} are $c_1^{BD}=1$ (independent of $\omega$), $c_2^{BD}=-0.400155$ when $\omega=500$ and $c_2^{BD}=-0.400232$ for $\omega=50000$ ($c_2^{BD}$ decreases with $\omega$). In the case of the supermassive Galactic black hole, using the same values of the parameters as above, if $\omega=500$ we obtain that $\theta_\infty^{BD}=26.4947$ $\mu$as, $s^{BD}=0.03316$ $\mu$as, $\mu_1^{BD}=6.4\times 10^{-13}$, $r^{BD}=r$, $|\Delta t^{\mathrm{o}\;\;\; BD}_{1,1}|=4.947 \times 10^{-10}$ min, and $|\Delta {t^{\mathrm{s}\;\;\; BD}_{1,2}}|=11.698$ min. If $\omega=50000$ we have that $\theta_\infty^{BD}=26.5091$ $\mu$as, $s^{BD}=0.03318$ $\mu$as, $\mu_1^{BD}=6.4\times 10^{-13}$, $r^{BD}=r$, $|\Delta t^{\mathrm{o}\;\;\; BD}_{1,1}|=4.950 \times 10^{-10}$ min, and $|\Delta {t^{\mathrm{s}\;\;\; BD}_{1,2}}|=11.704$ min. The values of all quantities are very close to the corresponding ones for the Schwarzschild black hole, and quite different from those for the phantom black holes with $b/m=3\pi/2$.

The observation of the vicinity of black holes will be possible in the next years, when new instruments are expected to be operational, in the radio and X bands, such as RADIOASTRON \cite{zakharov,webradio}, Event Horizon Telescope  \cite{eht} and MAXIM \cite{webmaxim}. RADIOASTRON is a space-based radio telescope, with an angular resolution of about $1-10 \, \mu \mathrm{as}$. The Event Horizon telescope is based on very long baseline interferometry, to combine existing and future millimeter/submillimeter facilities into a high-sensitivity, high angular resolution telescope. The MAXIM project is a space-based X-ray interferometer with an expected angular resolution of about $0.1 \, \mu \mathrm{as}$. Some observational features of the Galactic supermassive black hole, including strong deflection traits, can be found in the recent review \cite{mmg}. Subtle effects coming from the comparison of different black hole models, such as those arising from alternative theories, will surely require more advanced future instruments. 
\\

Note added: The day before this work was sent to arXiv, the paper \cite{ch} appeared
online in the same database, containing a partial overlap with our results.

\section*{Acknowledgments}

This work has been supported by CONICET. We thank an anonymous referee for very useful comments and suggestions.

\end{document}